\RequirePackage{latexrelease}
\documentclass[twocolumn,english,superscriptaddress,citeautoscript,preprintnumbers,amsmath,amssymb,prl,floatfix,footinbib]{revtex4}
\usepackage{hyperref}
\hypersetup{colorlinks=true,linkcolor=red,citecolor=blue}
\usepackage[scaled=2]{helvet}
\usepackage[T1]{fontenc}
\usepackage[latin9]{inputenc}
\usepackage{array}
\usepackage{float}
\usepackage{multirow}
\usepackage{amsmath}
\usepackage{amssymb}
\usepackage{graphicx}
\usepackage{csquotes}
\usepackage{xcolor,soul}

\makeatletter

\providecommand{\tabularnewline}{\\}

\@ifundefined{textcolor}{}
{%
	\definecolor{BLACK}{gray}{0}
	\definecolor{WHITE}{gray}{1}
	\definecolor{RED}{rgb}{1,0,0}
	\definecolor{GREEN}{rgb}{0,1,0}
	\definecolor{BLUE}{rgb}{0,0,1}
	\definecolor{CYAN}{cmyk}{1,0,0,0}
	\definecolor{MAGENTA}{cmyk}{0,1,0,0}
	\definecolor{YELLOW}{cmyk}{0,0,1,0}
}

\usepackage[scaled=2]{helvet}\usepackage{amstext}

\makeatletter

\usepackage{dcolumn}
\usepackage{bm}
\usepackage{times}

\makeatother

\usepackage{babel}

\makeatother
\usepackage{babel}
\begin{document}
	\title{Unconventional magneto-resistance, and electronic transition in Mn$_{3}$Ge Weyl
		semimetal}
	\author{V. Rai}
	\affiliation{Forschungszentrum J\"ulich GmbH, J\"ulich Centre for Neutron Science (JCNS-2)
		and Peter Gr\"unberg Institut (PGI-4), JARA-FIT, 52425 J\"ulich, Germany}
	\affiliation{RWTH Aachen, Lehrstuhl f\"ur Experimentalphysik IVc, J\"ulich-Aachen Research
		Alliance (JARA-FIT), 52074 Aachen, Germany}
	\author{S. Jana}
	\affiliation{Forschungszentrum J\"ulich GmbH, J\"ulich Centre for Neutron Science (JCNS-2)
		and Peter Gr\"unberg Institut (PGI-4), JARA-FIT, 52425 J\"ulich, Germany}
	\affiliation{RWTH Aachen, Lehrstuhl f\"ur Experimentalphysik IVc, J\"ulich-Aachen Research
		Alliance (JARA-FIT), 52074 Aachen, Germany}
	\author{M. Meven}
	\affiliation{J\"ulich Centre for Neutron Science at Heinz Maier-Leibnitz Zentrum,
		Forschungszentrum J\"ulich GmbH, Lichtenbergstra\ss{}e 1, 85747 Garching,
		Germany}
	\affiliation{Institute of Crystallography, RWTH Aachen University, 52056 Aachen,
		Germany}
	\author{R. Dutta}
	\affiliation{J\"ulich Centre for Neutron Science at Heinz Maier-Leibnitz Zentrum,
		Forschungszentrum J\"ulich GmbH, Lichtenbergstra\ss{}e 1, 85747 Garching,
		Germany}
	\affiliation{Institute of Crystallography, RWTH Aachen University, 52056 Aachen,
		Germany}
	\author{J. Per\ss{}on}
	\affiliation{Forschungszentrum J\"ulich GmbH, J\"ulich Centre for Neutron Science (JCNS-2)
		and Peter Gr\"unberg Institut (PGI-4), JARA-FIT, 52425 J\"ulich, Germany}
	\author{S. Nandi}
	\email{s.nandi@fz-juelich.de}	
	\affiliation{Forschungszentrum J\"ulich GmbH, J\"ulich Centre for Neutron Science (JCNS-2)
		and Peter Gr\"unberg Institut (PGI-4), JARA-FIT, 52425 J\"ulich, Germany}
	\affiliation{RWTH Aachen, Lehrstuhl f\"ur Experimentalphysik IVc, J\"ulich-Aachen Research
		Alliance (JARA-FIT), 52074 Aachen, Germany}
	\begin{abstract}
		Weyl semimetals are well known for their anomalous transport effects
		caused by a large fictitious magnetic field generated by the non-zero Berry curvature.  We performed the analysis of the electrical
		transport measurements of the magnetic Weyl semimetal Mn$_{3}$Ge
		in the \textit{a-b} and \textit{a-c} plane. We have observed negative longitudinal magneto-resistance (LMR) at a low magnetic field ($B<1.5$ T) along all the axes. The high field LMR shows different behavior along \textit{x} and \textit{z} axes. A similar trend has been observed in the case of planar Hall effect (PHE) measurements as well. The nature of high field LMR along the \textit{x} axis changes near 200 K. Dominant carrier concentration type, and metallic to semimetallic transition also occur near 200 K. These observations suggest two main conclusions: (i) The high field LMR in Mn$_3$Ge is driven by the metallic - semimetallic nature of the compound. (ii) Mn$_3$Ge compound goes through an electronic band topological transition near 200 K. Single crystal neutron diffraction does not show any change in the magnetic structure below 300 K. However, the in-plane lattice parameter (\emph{a}) shows a
		maximum near 230 K. Therefore, it is possible that the change in electronic band structure near 200 K is driven by the \textit{a} lattice parameter of the compound.
	\end{abstract}
	\maketitle
	
	\pagenumbering{arabic}
	\section{Introduction:}
	
	Topological materials have been extensively studied recently due to
	their observed anomalous transport effects (ATE). Signatures of Weyl fermions can be observed in topological materials, which exhibit broken inversion or time-reversal
	symmetry \cite{nayak2016large, nakatsuji2015large, armitage2018weyl, liu2018giant, xiong2015evidence}. Magnetic Weyl semimetals show large anomalous conductivity due to a
	strong Berry curvature \cite{shekhar2018anomalous,liu2018giant,nayak2016large,nakatsuji2015large}.
	Magnetic properties of Mn$_{3}$\textit{X} (\textit{X} = Sn, Ge, Ga)
	type of compounds were studied a long time ago \cite{tomiyoshi1982magnetic,tomiyoshi1983triangular,nagamiya1982triangular,kren1970neutron}.
	However, ATE in such compounds (\textit{X}
	= Ge, Sn) were observed only recently \cite{nayak2016large,nakatsuji2015large,liu2017transition,chen2014anomalous,zhang2016giant,kren1970neutron}.
	The hexagonal phase of these compounds has a non-collinear antiferromagnetic
	(AFM) structure in the \textit{a-b} plane, along with a small ferromagnetic
	(FM) moment which arises due to the spin canting away from the ideal
	triangular magnetic structure \cite{nayak2016large,tomiyoshi1983triangular,kiyohara2016giant}.
	The non-collinear AFM structure causes a non-vanishing Berry curvature,
	which gives rise to the ATE \cite{kiyohara2016giant,kubler2014non, jeon2021long}.
	The presence of a small in-plane FM moment in Mn$_{3}$\textit{X}
	helps in controlling the chirality of the magnetic structure within
	200 Oersted (Oe) of the external magnetic field. The observed large change in
	Hall resistivity in near-zero fields suggests that a large fictitious field is present in the system \cite{kiyohara2016giant, jeon2021long},
	which can be switched by 200 Oe of the external magnetic field. Therefore,
	compounds with large anomalous Hall effect (AHE) can be useful for developing spintronics
	devices.
	
	Among the family of Mn$_{3}$\textit{X}, the magnitude of anomalous
	hall conductivity in Mn$_{3}$Ge was predicted and observed to
	be the highest \cite{kubler2014non,nayak2016large,kiyohara2016giant}.
	Other than this, Mn$_{3}$Ge stands out because large anomalous transport
	effects were observed below 365 K, down to 2 K \cite{shekhar2018anomalous,liu2018giant,nayak2016large,nakatsuji2015large}.
	The observed anomalous Nernst effect \cite{hong2020large, wuttke2019berry}, AHE \cite{nayak2016large,kiyohara2016giant},
	and magneto-optical Kerr effect \cite{wu2020magneto} are signatures
	of the presence of the Weyl points in Mn$_{3}$Ge. The field dependent
	planar Hall effect has been reported recently \cite{xu2020planar}.
	However, chiral anomaly arises in Weyl semimetals due to the increase in conductivity primarily driven by the angle between the electric
	field (\textbf{\textit{E}}) and the magnetic field (\textbf{\textit{B}}).
	The correction term is proportional to \textbf{E}$\cdot$\textbf{B}. Therefore,
	angle-dependent electrical transport measurements can explicitly verify the presence
	of Weyl points in the system.
	
	Our primary aim in this paper is to interpret the origin of magneto-electrical transport effects. We performed magneto-resistance
	(MR) measurements with magnetic field and electric current (\textit{I}) applied
	along the \textit{x}, \textit{y}, and \textit{z} directions.  Here, \textit{x}, \textit{y} axes are along the
	{[}$2\bar1\bar10${]}, {[}$01\bar10${]}
	crystallographic directions, respectively, and both lie in the \textit{\emph{the}}\textit{
		a-b} plane of the hexagonal lattice. \textit{z} axis is along the {[}0001{]}
	direction, \textit{\emph{the}}\textit{ c} axis
	of the hexagonal lattice. Unlike Mn$_3$Sn {\cite{kuroda2017evidence}}, the LMR in Mn$_3$Ge is negative and positive at below and above $\sim$ 1.5 T. Therefore, the presence of chiral anomaly effect in Mn$_3$Ge cannot be justified. However, the sudden change in LMR near 1.5 T suggests a strong competition between two competing phenomena. We also report the observation of anisotropic MR and planar Hall effect (PHE) when the magnetic field and electric current
		are rotated in the \textit{ a-b} or \textit{a-c}
		plane. When a high magnetic field is applied along the \textit{x} axis, the slope of LMR (along the \textit{x} axis) remains positive and negative below and above 200 K, respectively. Such a change in the behavior of high field LMR can be linked with the metal to semimetal transition, and change in carrier concentration of the compound, near 200 K. These effects, including temperature dependent resistivity, indicate that the Mn$_{3}$Ge go through an electronic (topological) transition around 200 K, similar to the thin film-Mn$_3$Ge at 50 K {\cite{wang2021robust}}, possibly driven by the change in  \textit{a} lattice parameter of the compound.
	
	\section{Experimental methods}
	
\begin{figure}[h]
		\includegraphics[width=8cm]{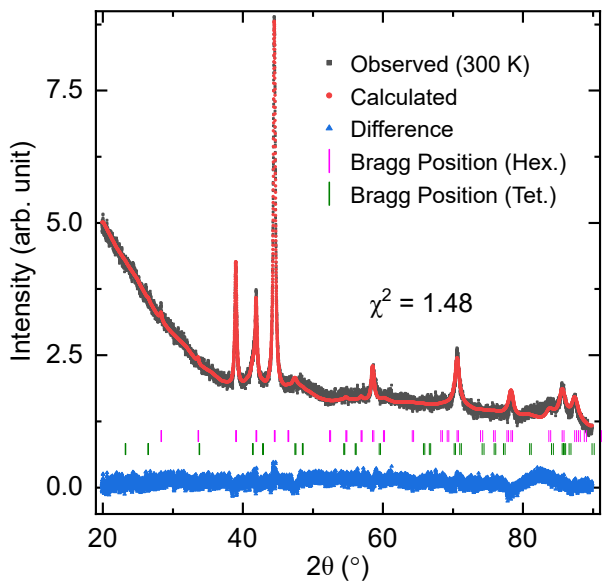} \caption{X-ray powder diffraction of Mn$_{3}$Ge at 300 K. Magenta and green
			color vertical lines show positions of Bragg peaks for tetragonal (Tet.)
			and hexagonal (Hex.) phases, respectively.}
		\label{fig:xrd}
	\end{figure}

	The experiments were performed on single crystals of ($\epsilon$)-Mn$_{3}$Ge (hexagonal
	phase), which is stable above 900 K,
	and metastable below. The hexagonal phase of Mn$_{3}$Ge
	forms only with excess in Mn. According to Refs. \cite{berche2014thermodynamic, binary1, yamada1988magnetic}, Hexagonal phase is stable only when Mn:Ge = $(3+\delta)$:$1$, where $\delta\approx0.15-0.60$. To prepare the Mn$_{3}$Ge single
	crystals, high purity ($>99.99\%$) Mn and Ge were taken in a stoichiometric
	ratio of $(3+\delta)$:$1$ ($\delta=0.1,0.2,0.5$), and then melted
	using induction melting technique to ensure the homogeneous mixing
	of the elements. After that, the alloy was sealed in a quartz tube
	and heated to 1273 K for 10 hours, followed by cooling at
	a rate of 2 K/\,hr down to 1073 K. Finally,
	the sample was quenched at 1073 K to retain the high-temperature
	hexagonal phase. The chemical analysis of the single crystals was
	done by using the ICP-OES method. The chemical compositions of the
	three single crystals were found to be Mn$_{3.08(5)}$Ge, Mn$_{3.18(5)}$Ge, and Mn$_{3.55(5)}$Ge, which will be referred ahead as S1, S2, and S3, respectively. For convenience, Mn$_{3+\delta}$Ge
	is termed in this paper as Mn$_{3}$Ge unless specified otherwise.
	The X-ray powder diffraction of the sample was performed by crushing
	a few single crystals from the same batch. Data analysis was performed
	using the FullProf software (Fig. \ref{fig:xrd}). The data analysis using the tetragonal and hexagonal phases of Mn$_3$Ge showed that crystals
	were synthesized mostly (96(2)\%) in the hexagonal phase with P$6_{3}$/\textit{mmc}
	space group symmetry. The lattice parameters at room temperature were
	found to be \textit{a} = \textit{b} = 5.3325(5) {\AA}, \textit{c} = 4.3122(4) {\AA}.
	Temperature-dependent X-ray diffraction was also performed using a 
	Huber Imaging Plate Guinier Camera G670 (Fig. \ref{fig:xrd_t}).
	
	Single crystal neutron diffraction measurements
	were performed on the (S3) sample, using the HEiDi instrument
	at the FRM II neutron source, Garching (Germany) (see section {\ref{neutron-sec.}}). The data analysis (using JANA2006 {\cite{JANA2006}}) has shown that the chemical composition of the sample is Mn$_{[3+0.09(1)]}$Ge$_{[1-0.09(1)]}$, which, after normalization, corresponds to Mn:Ge = 3.40(5):1, slightly lower than the chemical composition of the same sample determined by the ICP-OES method. Neutron diffraction analysis also confirms that the excess  $\sim$10\% Ge sites are occupied by the Mn atoms. Further details of the data analysis will be discussed in section {\ref{neutron-sec.}}.
	
	Electrical transport and magnetization measurements were performed on the three samples (S1, S2, S3) using the Quantum Design Physical Property Measurement System (PPMS).	However, most of the transport measurements were performed using	S1 and S2 samples because of having similar chemical compositions
	as reported in the literature \cite{wuttke2019berry,nayak2016large,kiyohara2016giant,chen2020antichiral,xu2020finite}. In some cases, sample S3 was also measured to determine the evolution of physical properties with the Mn concentration. Our analysis remained consistent with all three samples.

	\section{Electrical Transport Results}
	
	\subsection{Longitudinal magneto-resistance} \label{LMR}
	We measured the longitudinal magneto-resistance (LMR) of Mn$_3$Ge for different combinations of the applied current and magnetic field directions. The measurements were repeated using different samples pieces with thin rectangular shapes (typical dimension - length: 1.5 - 2.1 mm; width: 0.4 - 0.5 mm; thickness: 0.1 - 0.3 mm).	 We observed a	consistent behavior of LMR above 0.05 T field in all three samples. The MR\% (=$100\times$$\frac{\rho(B)-\rho(0)}{\rho(0)}$) of all the samples lies in the range of 0.5 - 1\%. Therefore, careful analysis is required to determine its origin. Even a small misalignment of the sample contacts led to an asymmetric
	curve in $\pm$\textit{B} regimes due to a large anomalous Hall contribution
	near zero field. Therefore, the MR data were symmetrized to
	extract the true MR contribution. Negative MR was observed in the low field
	regime where I$\parallel$\textit{B}$\parallel$(\textit{x}, \textit{y} and \textit{z})
	axes (Fig. \ref{fig:mr_para_perp} in the Appendix B). The nature of LMR was found
	to be very similar as long as the magnetic field and current are applied
	in the \textit{x-y} (or \textit{a-b}) plane. However, it is different when the field and current are applied along the
	\textit{z} axis. The LMR data show non-monotonic behavior with the magnetic field in all the cases. Most of the electric transport measurement analysis presented here corresponds to the S2 sample unless another sample is mentioned. 
	
	\begin{figure*}
		\includegraphics[width=17cm]{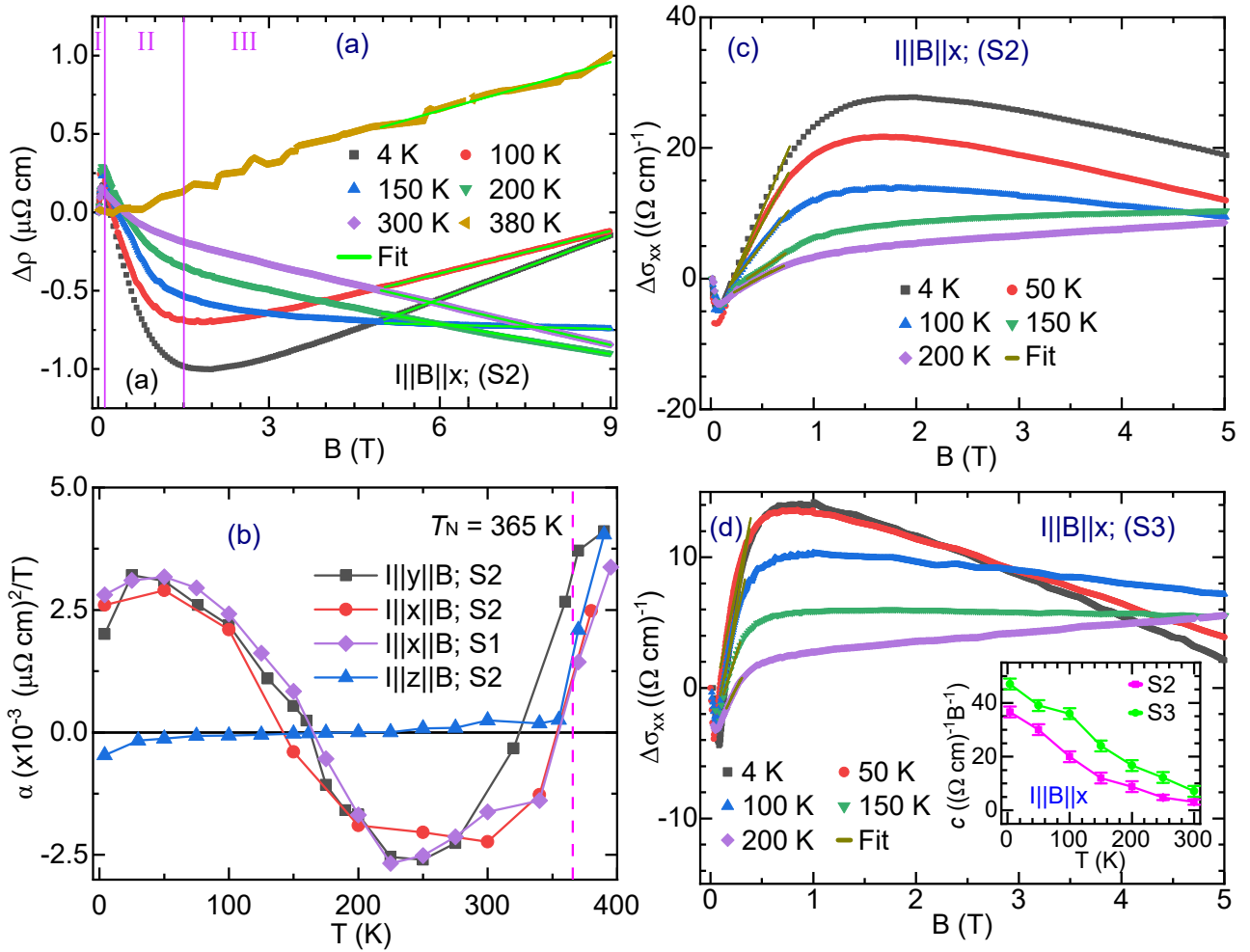}\\
		\caption{(a) LMR of Mn$_{3}$Ge (S2) at different temperatures with I$\parallel$\textit{B}$\parallel$\textit{x}. Here, $\Delta\rho = \rho(B)-\rho(0)$. At high field (5 T - 9 T), the data is fitted (Fit) linearly ($\rho_{[\text{high }B]}\approx\alpha (B/\rho_0)$) to determine the slope - $\alpha$. I, II, and III denote the different field regimes where the nature of LMR is different. (b) Temperature dependence of high field slope, $\alpha$, corresponding to the different samples. The nature of $\alpha$ is almost the same for	all samples as long as  the magnetic field lies in
			the \textit{x-y} plane. However, it is different when \textit{B}$\parallel$\textit{z}
			is applied. A change of the sign of ($\alpha$) occurs near 165 K in all the cases. (c, d) show LMC corresponding to the S2 and S3 samples, respectively, with field applied along the \textit{x} axis. $\Delta\sigma_{xx}= \sigma_{xx}(B)-\sigma_{xx}(0)$.  At low field, positive $\Delta\sigma_{xx}$ fits linearly with field ($\sigma_{[\text{low }B]}\approx cB$). The temperature dependence of the slope, $c$, corresponding to both the samples is shown in the (d) inset.}
		\label{fig:LMR_LMC_all}
	\end{figure*}

	The LMR along \textit{x} axis, for S2 and S3 samples, increases with the magnetic field up to $\sim$ 0.05 T, which is referred to as region I,  as shown in Fig. {\ref{fig:LMR_LMC_all}}(a). However, such behavior was not observed when the field was applied along the \textit{y}, \textit{z} axes (Fig. {\ref{fig:mr_para_perp}} in the appendix). Such an increase in LMR at a very low field was not observed when the LMR of the S1 sample was measured along the \textit{x} axis (Fig. {\ref{fig:mr_side}} in the appendix). Beyond 0.05 T, LMR starts to show a sudden decrease till $\sim$ 1.5 T (S2 sample), referred as region II. This behavior was observed in all cases, irrespective of the sample, and the direction of the magnetic field. Near 1.5 T, negative MR starts to get suppressed, followed by region III. The LMR in region III shows different behavior in different cases, which will be discussed later. In the case of the S3 sample, a similar change from region II to region III is observed near 0.8 T. Similar nature of LMR has also been observed in the case of the thin-film Mn$_3$Ge {\cite{qin2020anomalous}}.

	As described and observed by Ref. \mbox{\cite{liang2018experimental, zhang2016signatures}}, the current jetting effect can lead to negative MR in anisotropic samples. Therefore, the effect of current jetting on the MR of our sample has been measured, as described in Appendix B. It is clear from the observation that the role of current jetting is insignificant in our samples. This is expected because in contrast with the Na$_3$Bi and GdPtBi {\cite{liang2018experimental}}, the magnitude of resistivity and MR in Mn$_3$Ge is nearly same along different axes (see Fig ({\ref{fig:rt}}, {\ref{fig:mr_para_perp}}) in the appendix). The absence of the current jetting effect provides a piece of strong evidence that the observed LMR is intrinsic.

	\subsection*{ \textit{I}$\parallel$\textit{B}$\parallel$\textit{x}}
	
	In the case of (\textit{I}$\parallel$\textit{B}$\parallel$\textit{x}), the
	origin of the region I is likely to be a weak antilocalization (WAL)
	effect. Its magnitude was found to be $<$0.05 $\mu\Omega$ cm,  0.3 $\mu\Omega$ cm, and 0.6 $\mu\Omega$ cm, in the case of S1, S2,
	and S3 samples, respectively. The increase in the magnitude of MR in the region I could link with the increase in the strength of the WAL	with the disorder in the system with an increase in the Mn concentration.

	The sudden decrease of LMR in region II is observed along all the axes, in all three samples. We have compared LMR in this field regime in terms of Longitudinal magneto-conductivity (LMC).	The relation between LMC ($\sigma_{ii}$) and LMR ($\rho_{ii}$) is given as: $\sigma_{ii}=\rho_{ii}^{-1}$. LMC for S2 and S3 samples are compared in Figs. {\ref{fig:LMR_LMC_all}}(c, d), where it is clear that the LMC increases almost linearly (beyond 0.05 T) up to 0.8 T and 0.4 T, in S2 and S3 samples, respectively. As mentioned above, this magnetic field regime is referred as region II.
	The LMC in region II shows positive (linear) increase when the magnetic field is applied along the \textit{x} axis, and can be fitted linearly (at low field) using the equation: $\sigma_{[\text{low }B]}\approx cB$. Where $c$ is the proportionality constant.
	
	The LMC along the \textit{x} axis shows positive and increasing behavior at low field. However, it starts to decrease for $B>1.5$ T. This observation is different from the Mn$_3$Sn, where positive LMC persists up to 9 T {\cite{kuroda2017evidence}}. Therefore, the presence of the chiral anomaly effect in Mn$_3$Ge cannot be justified.

	As shown in  Fig. {\ref{fig:LMR_LMC_all}(d)}, $c$ at 200 K is nearly one-third of with $c$ at 4 K. However, the magnetization (Fig. {\ref{fig:mt}} in the appendix) remains almost constant up to $\sim$ 200 K. Furthermore, the field dependent magnetization is linear, however, LMC is not monotonic, up to 3 T, with the magnetic field. These observations imply that the positive LMC in region II is not driven by the magnetization of the sample.
	
	At high magnetic field (region III), LMR shows linear field-dependent behavior with slope $\alpha$ $($$\rho_{[\text{high }B]}\approx\alpha (B/\rho_0)$$)$, where $\rho_0$ is the zero field resistivity. The temperature dependence of $\alpha$ for three samples with field applied along different axes is shown in Fig. {\ref{fig:LMR_LMC_all}}(b). Unlike Mn$_3$Sn {\cite{kuroda2017evidence}}, positive LMR ($\alpha>0$) is observed in Mn$_3$Ge below $\sim$ 165 K when field is applied along the \textit{x, y} axes. Positive LMR has been observed in several Weyl semimetals \mbox{\cite{ghimire2018anisotropic, sharma2020observation,zhang2016signatures, huang2015observation, zhang2020magnetotransport, li2017negative}}, however, such a feature has been attributed to the extrinsic effects, like current jetting or misalignments {\cite{li2017negative}}. Since we have already nullified the possibility of the current jetting effect, the nature and origin of $\alpha$ have to be intrinsic. Application of the magnetic field in a different direction can lead to the opening of the gap in the Weyl points, or Dirac points, which can lead change in the nature of MR {\cite{zeng2020intrinsic}}. Such electronic transition has been observed in TaAs (Weyl semimetal), where vanishing Weyl points lead AHE to vanish near 50 T {\cite{ramshaw2018quantum}}. However, in the case of Mn$_3$Ge, AHE is present up to 9 T, which suggests that the Weyl points remain separated beyond 1.5 T as well. Therefore, change in LMR near 1.5 T does not originate due to field induced electronic transition.

	In the case of Mn$_3$Ge, metallic to non-metallic transition is observed in the resistivity measurements (see Fig. {\ref{fig:rt}} in the appendix) near 200 K. We assume that the non-metallic region (above 200 K) is semimetal. However, spin-resolved band structure calculations are required to determine whether it is a semimetal or half-metal above 200 K. For convenience, we will consider our sample at $T>200$ as semimetal. 
	Positive and negative longitudinal MR can be observed in the case of metallic, and halfmetallic/semimetallic samples \mbox{\cite{yang2012anisotropic,  pippard1989magnetoresistance, roth1963empirical, furukawa1962magnetoresistance, kawabata1980theory, Endo1999MagnetoresistanceOC, sun2020room}}, respectively, depending on the spin polarized density of states (DOS) near the Fermi surface {\cite{kokado2012anisotropic}}. Crossover of MR from positive to negative MR due to change in the spin polarized DOS, near the Fermi surface, has been reported in the case of Half metallic compounds \mbox{\cite{ouardi2013realization, du2013crossover}}.
	Since Mn$_3$Ge shows semimetal to metal transition near 200 K, the high field LMR at low and high temperatures possibly originates due to the metallic and semimetallic nature of the sample, respectively.
	At lower temperatures, low field LMR is negative, and high field LMR is positive. This suggests a strong competition between two different phenomena, which leads to a change in sign of $\alpha$ near 165 K, rather than near 200 K, where semimetal to metal transition occurs. It is important to note that the high field positive LMR due to the metallic nature of the sample has not been observed in the case of Mn$_3$Sn \mbox{\cite{kuroda2017evidence, chen2021anomalous}}. Such difference in LMR for Mn$_3$Ge and Mn$_3$Sn is possible because electrical transport measurements are intimately linked with the DOS, which differs significantly for these compounds, as reported by Refs. \mbox{\cite{yang2017topological, kubler2014non}}.
	
	Similar observation of crossover of MR, at an intermediate temperature, due to the relative change in the spin polarized DOS, has been observed in the case of ferromagnetic thin-film CrTe$_2$, Half metallic, and semimetallic compounds as well \mbox{\cite{sun2020room, du2013crossover, ouardi2013realization}}. Therefore, it would be very interesting to study the spin polarized DOS in Mn$_3$Ge at low and high temperatures to justify the role of $3d$ spins in the observed LMR. 
	
	\begin{figure}
		\includegraphics[width=8cm]{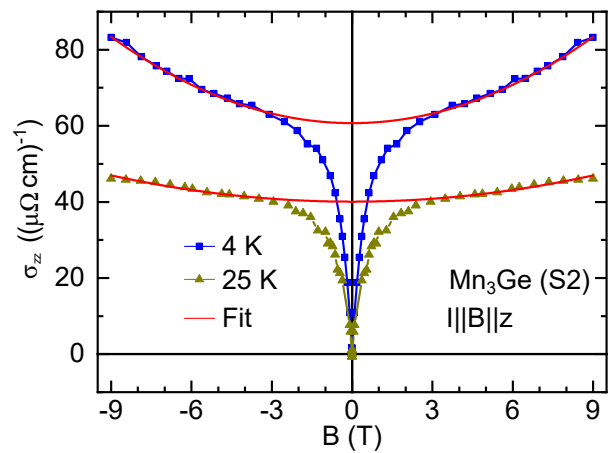} 
		\caption{Longitudinal magneto-conductivity of the Mn$_3$Ge (S2) with the magnetic and electric current applied along the \textit{z} axis. The data is fitted with $a_1+a_2B^2$, where $a_1$ and $a_2$ denote the fitting parameters.}
		\label{fig:LMC_zz}
	\end{figure}

	\subsection*{ \textit{I}$\parallel$\textit{B}$\parallel$\textit{z}}
	
	In contrast to the \textit{x} axis, LMC along the \textit{z} axis increases with the magnetic field up to 9 T, as shown in Figs. {\ref{fig:LMC_zz}} (and Fig.  {\ref{fig:LMC_B_IzzBz}} in the appendix).
		The sharp linear increase in LMC, along \textit{z} axis, at the low field is possibly due to the domain effect, as suggested by Ref. {\cite{chen2021anomalous}}.
		For $B>3$ T, we have observed $B^2$ dependence of the LMC, at 4 K, similar to the  Ref. {\cite{chen2021anomalous}}. However, such a field dependence, and magnitude of LMC, both weaken as the temperature is increased to 25 K.

	\subsection{Angle dependent MR ($\theta$MR)} 
	\label{thetaMR}
	
	\begin{figure}
		\includegraphics[width=8cm]{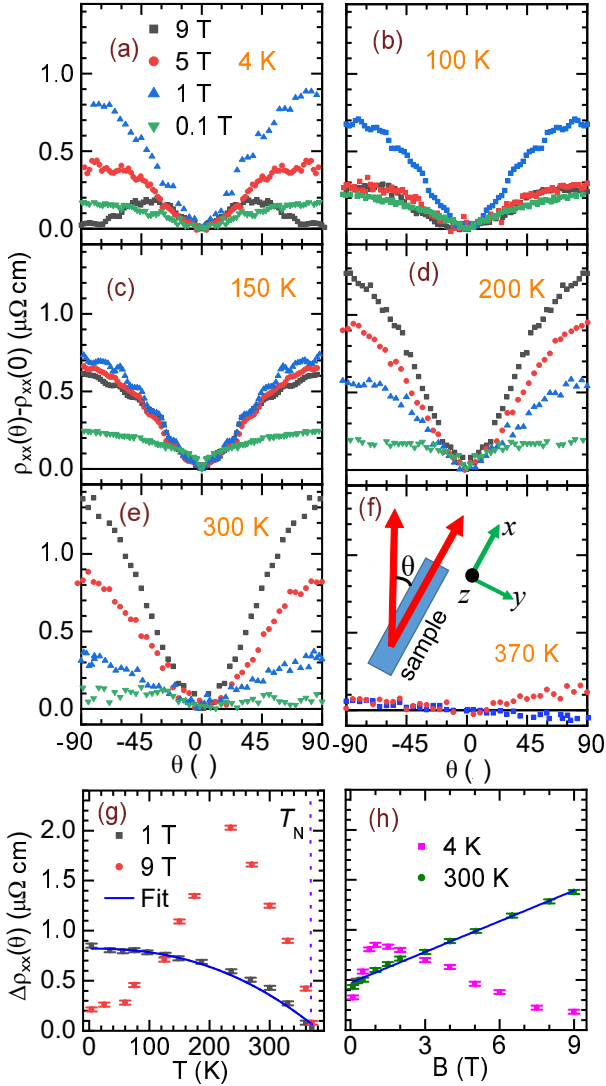} \caption{ (a-f) Angle dependent MR at different magnetic fields and temperatures. The experimental setup for $\theta$MR measurement is shown in the (f). Where, the current is applied along the \textit{x} axis, and the sample is rotated towards the \textit{y} axis, making an angle $\theta$ with the magnetic field direction. (g, h) Temperature and field dependence of the magnitude of the $\theta$MR oscillations are shown above. Here, $\Delta\rho_{xx}=\rho(\theta_{max})-\rho(0)$.  Where $\theta_{\text{max.}}$ denotes the angle at which $\theta$MR is observed to be maximum. Usually, $\theta_{\text{max.}}=90^{\circ}$, except for low temperature measurements performed at high field. In (g, h),  \enquote*{Fit} denotes the fitting with $a-bT^m$, and $B^n$, respectively, as mentioned in the text. The fitting parameters, $m=2.3(2)$ and $n=1.00(5)$ were obtained.}  
		\label{fig:mr_th_xy}
	\end{figure}

	Angle dependent MR (referred as $\theta$MR) was performed by rotating the sample, thus making the current	direction have an angle $\theta$ with respect to the applied magnetic field. All the $\theta$MR measurements were performed using the S2 sample, whose corresponding LMR is shown in Fig. \ref{fig:LMR_LMC_all}(a). The $\theta$MR measurement can be performed in several different
	combinations of the direction of rotation, magnetic field, and electric
	current. Therefore, we performed $\theta$MR in four different
	ways: (i) \textit{I}$\parallel$\textit{x}; \textit{B} is rotation from $x\rightarrow$\textit{y}
	axis, (ii) \textit{I}$\parallel$\textit{y}; \textit{B} is rotation from $y\rightarrow$\textit{x}
	axis, (iii) \textit{I}$\parallel$\textit{x}; \textit{B} is rotation from $x\rightarrow$\textit{z}
	axis, (iv) \textit{I}$\parallel$\textit{z}; \textit{B} is rotation from $z\rightarrow$\textit{x}
	axis. Field dependent behavior of $\theta$MR is different depending on the direction of the current at $\theta=0^{\circ}$, and the plane of rotation. $\theta$MR remains nearly same (except at \textit{B} = 9 T at 4 K) as long as the field is rotated in the \textit{x}-\textit{y} plane, which is the case (i, ii) as shown in Figs. (\ref{fig:mr_th_xy}, \ref{fig:mr_th_yx_zx}(a - c) in the appendix). However, it is different in other two cases when plane of rotation is \textit{x}-\textit{z} axis (compare Figs. \ref{fig:mr_th_yx_zx}(d, e, f), and Fig. \ref{fig:mr_th_xz} in the appendix). 
	
	The MR of the sample with \textit{B} applied parallel ($\rho_{\parallel}$) and perpendicular ($\rho_{\bot}$) to the current direction axis is compared in Fig. {\ref{fig:mr_para_perp}} in the appendix. Based on this, it is obvious that the $\theta$MR, ($=\rho_{\bot}-\rho_{\parallel}$) has to be positive in almost all the cases, except high field measurements at 4 K, and the same has been observed during the $\theta$MR measurements.

	Positive $\theta$MR is observed when it is measured corresponding to  \textit{I}$\parallel$\textit{B}$\parallel$\textit{x}; \textit{B} rotation from $x$ $\rightarrow$$y$ (Figs. {\ref{fig:mr_th_xy}}(a-f)). The magnitude and behavior of $\theta$MR behaves in a similar way when it is measured for \textit{I}$\parallel$\textit{B}$\parallel$\textit{y}; \textit{B} rotated from $y$ $\rightarrow$$x$ (see Fig. {\ref{fig:mr_th_yx_zx}}(a-c) in the appendix).
	We have performed detailed analysis of $\theta$MR corresponding to \textit{I}$\parallel$$\textit{B}$$\parallel$\textit{x}; \textit{B} rotation from $x$ $\rightarrow$$y$. The temperature dependence of magnitude of $\theta$MR is shown in Fig. {\ref{fig:mr_th_xy}}(g). It has been observed that the magnitude of $\theta$MR at 9 T increases with temperature up to $\sim$ 200 K, beyond which it starts to decrease and vanishes near 365 K (\textit{T}$_{\text{N}}$). 
	The $\theta$MR at 9 T is unlikely to be originated from the chiral anomaly effect, because
	the LMR increases with field beyond 1.5 T at 4 K (Fig. {\ref{fig:LMR_LMC_all}}(a)). In contrast with 9 T, $\theta$MR at 1 T monotonically decreases with temperature, and can be fitted with  $a-bT^{\text{m}}$, where m = 2.3(2) was observed, which is higher than observed in other Dirac or Weyl semimetals (m = 1.4 {\cite{singha2018planar}}, m =1.7 {\cite{sharma2020observation}}). The temperature dependence of $\theta$MR magnitude at 1 T is similar to the temperature dependent magnetization Fig. ({\ref{fig:mt}}). Therefore, low field $\theta$MR might originate from the tiny magnetization of the sample.
	
	As shown in Fig. {\ref{fig:mr_th_xy}}(h), at low temperature (4 K), the magnitude of $\theta$MR increases with field up to 1.5 T, after which it starts to decrease up to 9 T.  Such a behavior is expected on the basis of LMR (Fig. {\ref{fig:LMR_LMC_all}}(a)) and transverse MR (Fig. {\ref{fig:mr_para_perp}} in the appendix) as well.

	\subsection*{Analysis of high field $\theta$MR}

	According to the  Refs. \mbox{\cite{kokado2012anisotropic, mcguire1975anisotropic}}, unequal $3d$ spin DOS near Fermi surface can lead to the positive and negative $\theta$MR, as observed by Refs. \mbox{\cite{sun2020room, yang2012anisotropic}}. The condition for negative and positive $\theta$MR$_{spin}$ is discussed in more detail in Ref. {\cite{kokado2012anisotropic}}. We will refer to the $\theta$MR contribution due to spin DOS as $\theta$MR$_{spin}$. The temperature dependent resistivity in Fig. {\ref{fig:rt}} (in the appendix) suggests that Mn$_3$Ge is metallic below $\sim$ 200 K, and semimetallic above this temperature, which suggests a possible change in the spin DOS, near the Fermi surface, near 200 K. Since our sample is metallic below 200 K, $\theta$MR$_{spin}$ should be negative  (i.e. $\rho_{\bot}$$<$$\rho_{\parallel}$)  
	in behavior \mbox{\cite{yang2012anisotropic, kokado2012anisotropic}}. As observed in Fig. {\ref{fig:LMR_LMC_all}}, high field LMR increases with the magnetic field, but it remains negative in magnitude due to its sharp decrease at low field (below 1.5 T). Also, the transverse MR (Fig. {\ref{fig:mr_para_perp}} in the appendix) is larger than LMR up to $\sim$ 8 T. Therefore, positive $\theta$MR ($\rho_{\bot}$$>$$\rho_{\parallel}$) is observed at 4 K, even though the sample is metallic at this temperature.	
	
	The field dependent $\theta$MR magnitude in Fig. {\ref{fig:mr_th_xy}}(h) shows a decrease with an increase in the magnetic field, which also suggest that high field $\theta$MR$_{spin}$ (expected to show negative $\theta$MR) suppresses the low field $\theta$MR (which leads to positive $\theta$MR), as the magnetic field is increased. These observations also suggest that similar to the high field LMR, the high field $\theta$MR is also driven by the spin DOS near the Fermi surface. Furthermore, the observed $\theta$MR at 4 K is a combined effect of two competing phenomena, which results in negative and positive LMR, below 200 K. Therefore, an increase in the magnitude of $\theta$MR, at 9 T, with temperature, up to $\sim$ 200 K, is possible if both the effects weaken with an increase in temperature at different rates  (see Fig. {\ref{fig:mr_th_xy}}(g)).

	Above $\sim$ 200 K, Mn$_3$Ge is semimetallic, which suggest that for $\theta$MR driven by unequal spin density ($\theta$MR$_{spin}$), has to be positive  $\rho_{\bot}$$<$$\rho_{\parallel}$ \mbox{\cite{yang2012anisotropic, kokado2012anisotropic}}. We have observed positive $\theta$MR at 300 K. The $\theta$MR at 300 K also increases linearly with an increase in the magnetic field (Fig. {\ref{fig:mr_th_xy}}(h)). This confirms that the positive $\theta$MR for \textit{T} $\geq$ 200 K (Fig. {\ref{fig:mr_th_xy}}(d, e)) originates due to the $\theta$MR$_{spin}$.

	\subsection{Anomalous Hall effect and planar Hall effect} \label{IIIc}
	
	\subsection*{Anomalous Hall effect}
	
	\begin{figure*}
		\includegraphics[width=17cm]{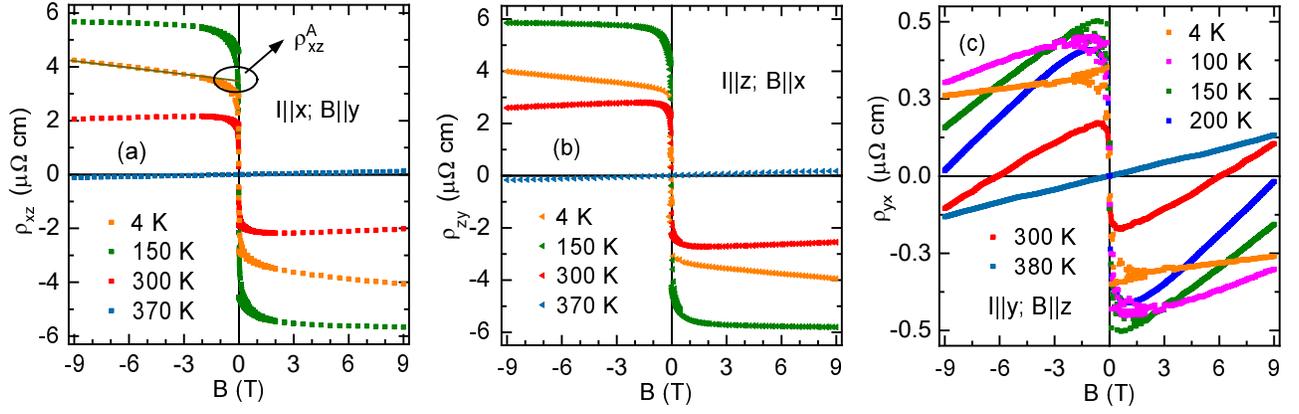} \caption{(a - c) shows Hall resistivity of Mn$_{3}$Ge (S2) with magnetic field applied along\textit{
			}\textit{\emph{the}}\textit{ y}, \textit{x} and \textit{z} axes, respectively. In case of
			(a), 4 K data is linearly fitted in the high field (-9
			T to -3 T) regime. The intercept of the fitted line, $\rho_{xz}^{\text{A}}$,
			is considered anomalous Hall resistivity, which is used to calculate
			the corresponding Hall conductivity. Such fitting was performed at several temperatures,
			and temperature dependent Hall conductivity along different axes was
			determined.}
		\label{fig:hall_all_axis}
	\end{figure*}

\begin{figure}
		\includegraphics[width=8.5cm]{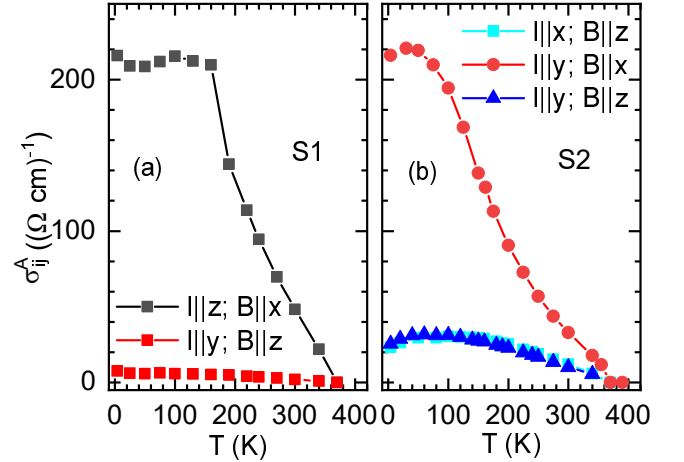} \caption{Temperature dependent anomalous Hall conductivity for (a) S1, (b) S2 sample, respectively, under different current and field direction.}
		\label{fig:cond_t}
	\end{figure}

	Hall resistivity of Mn$_{3}$Ge was measured in different combinations of the current and magnetic field. The measurements were repeated using different samples as mentioned previously (section \ref{LMR}).	Consistent behavior of Hall resistivity was observed during all the measurements performed using different samples. Fig. \ref{fig:hall_all_axis} shows Hall resistivity of Mn$_{3}$Ge (S2) with field applied along the \textit{y}, \textit{x} and \textit{z}
	axes at various temperatures. It can be seen that Hall resistivity
	shows a finite jump at zero magnetic fields followed by a linear field
	dependence at higher fields. The Hall effect can arise because of the
	external or internal magnetic field present in the system. In the
	case of Weyl semimetals, Berry curvature over the Brillouin zone does
	not vanish, which results in a huge fictitious magnetic field, giving
	rise to the AHE. Hall resistivity ($\rho_{ij}$), with current and voltage measured along $i$ and \textit{j} axis, can be written into its components
	as: $\rho_{ij} =R_{B}B+R_{s}M+\rho_{ij}^{A}$. Where, the first term
	shows normal Hall resistivity contribution, which is proportional to the external magnetic
	field (\textit{B}). $R_{s}M$ is the Hall resistivity component that originates
	due to the magnetization (\textit{M}) of the sample. The term $\rho_{ij}^{A}$ denotes the anomalous Hall resistivity due to the non-zero Berry curvature. In the case of Mn$_{3}$Ge,
	a small residual magnetic moment is present but the $\rho_{H}$-$H$
	the curve does not follow the \textit{M-H} type behavior. Also, the temperature-dependent
	Hall conductivity does not follow the \textit{M-T}
	behavior (compare Fig. \ref{fig:cond_t} and Fig. \ref{fig:mt}). Therefore, we conclude
	that the magnetization plays a negligible role in the anomalous Hall
	resistivity observed in Mn$_{3}$Ge \cite{kiyohara2016giant} and
	most likely originates from the non-zero Berry curvature, which is denoted by $\rho_{ij}^{A}$.
	
\begin{figure}
		\includegraphics[width=8cm]{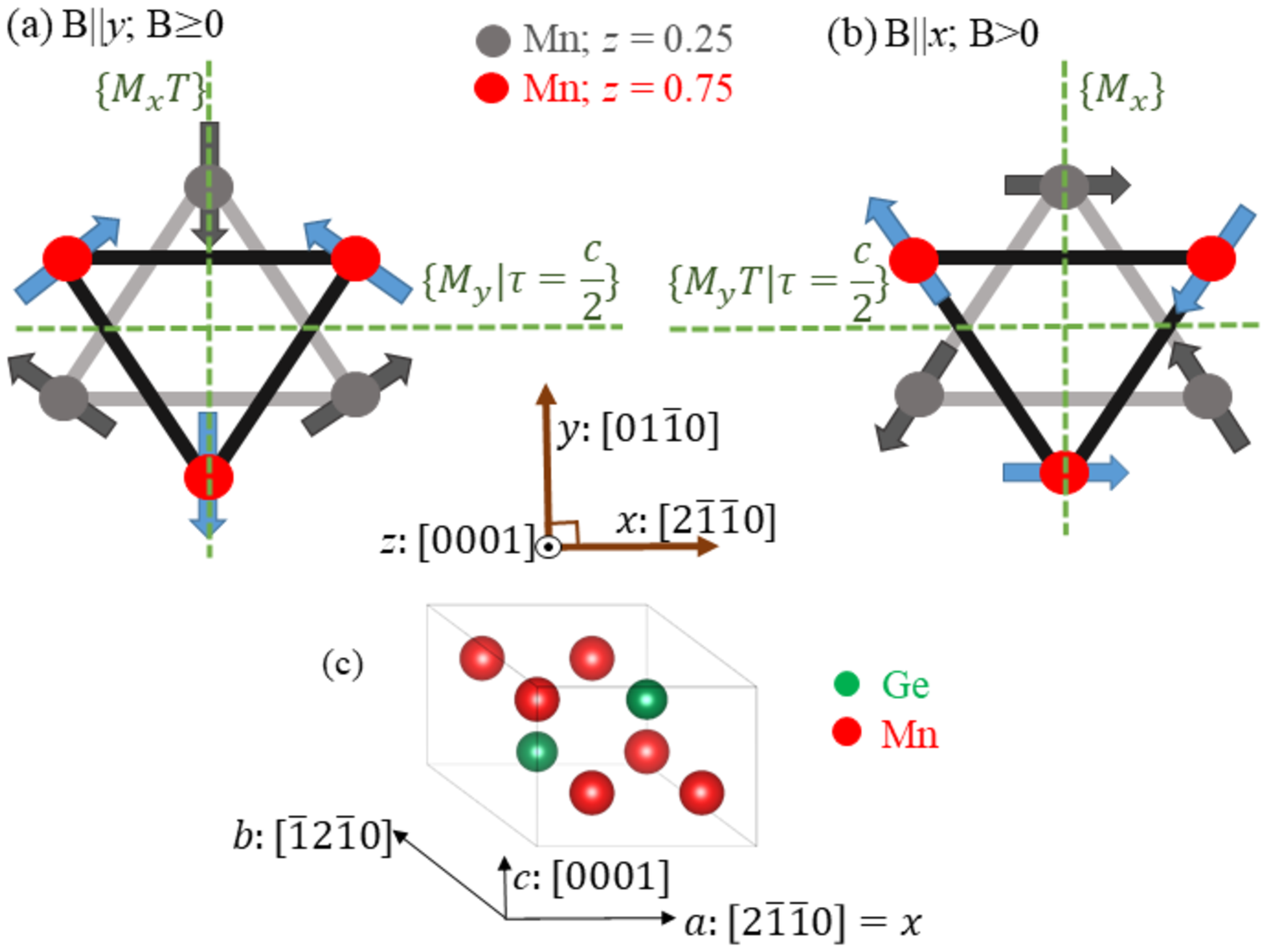} \caption{(a) Magnetic structure of Mn$_{3}$Ge in absence of a magnetic field or if the field applied along the \textit{y} {[}$0\thinspace1\bar{\thinspace1\thinspace}0${]}
			axis. (b) The magnetic structure when the magnetic field is applied along
			\textit{x} {[}$2\bar{\thinspace1}\bar{\thinspace1}\thinspace0${]}
			axis. \cite{tomiyoshi1983triangular,nagamiya1982triangular}. Mn atoms
			form Kagome type lattice structure with moments oriented by 120$^{o}$
			relative to each other. Gray and red Mn layers are located at \textit{z}
			= 0 and \textit{z} = 1/2 positions, respectively. Dashed green lines
			shown symmetry planes, where M$_{x}$, M$_{y}$, shows mirror plane
			in \textit{x-z} and \textit{y-z} planes. \textit{T} denotes the time reversal
			symmetry, and $\tau=\frac{c}{2}$ denotes the translation along \textit{c} axis by $\frac{c}{2}$. (c) Crystal structure of the Mn$_{3}$Ge.}
		\label{fig:mag_define}
	\end{figure}

	The non-zero value of the anomalous Hall resistivity is intimately connected to the underlying symmetry of the magnetic structure. The ideal magnetic structure
	of Mn$_{3}$Ge is a coplanar triangular antiferromagnetic structure with spins
	aligned in a reverse triangular arrangement (Fig. \ref{fig:mag_define}(a)).
	However, the presence of Dzyaloshinskii-Moriya interaction (DMI) in such a system, leads to small ferromagnetic canting, which has two important consequences: (i) The presence of only a few hundred Oersted (Oe) magnetic fields can create unequal domain population of spins with opposite chirality, which leads to the non-zero value of Berry curvature. 
	(ii) The Mn spin triangle can be rotated easily with the help of just a few hundred Oe of magnetic
	field \cite{tomiyoshi1983triangular, nagamiya1982triangular}. Therefore, once the magnetic field is reversed, Hall
	resistance also reverses its sign within 200 Oe of the applied
	magnetic field (Fig. \ref{fig:hall_hyst}).
	
	The symmetry analysis performed by Ref. \cite{nayak2016large}
	claims that $\sigma_{zx}^{A}$ (\textit{B}$\parallel$\textit{y})
	shows large intrinsic Hall conductivity. However, $\sigma_{xy}^{A}$
	(\textit{B}$\parallel$\textit{z}), $\sigma_{zy}^{A}$ (\textit{B}$\parallel$\textit{x})
	should vanish if the zero-field magnetic structure is taken into consideration.
	Here, $\sigma{_{ij}^{A}}$ denotes that the current (voltage) is applied
	(measured) along the \textit{i} (\textit{j}) axis, and the magnetic field
	is perpendicular to both directions. Despite their claim, a large
	and comparable magnitude of anomalous Hall conductivity (AHC) was observed
	if the magnetic field was applied along the \textit{x} and \textit{y}
	axis \cite{nayak2016large}. We also observed large anomalous
	Hall resistivity, which corresponds to large AHC, when the magnetic field was applied along the \textit{x}
	or \textit{y} axis (Fig \ref{fig:hall_all_axis}). The presence of a large
	$\sigma_{zy}^{A}$ can be explained by the field induced change of
	the magnetic structure of Mn$_{3}$Ge.
	
	Based on field-dependent single crystal neutron diffraction, Refs. \cite{tomiyoshi1983triangular,nagamiya1982triangular}
	claim that the magnetic structure of Mn$_{3}$Ge
	changes if a magnetic field is applied along the \textit{x} or \textit{y} axis. Fig. \ref{fig:mag_define}(a)
	shows the ground state magnetic structure, which remains the same even if the magnetic field is applied along the \textit{y} axis. However, if the field is applied along the \textit{x} axis, the magnetic structure changes as shown in Fig. \ref{fig:mag_define}(b). In the case of \textit{B}$\parallel$\textit{y}, and \textit{B}$\parallel$\textit{x}, there exists a mirror plane ($M_y$, and $M_x$, respectively) as shown in Fig. \ref{fig:mag_define}(a, b). Mirror reflection reverses the chirality but keeps the Berry curvature preserved \cite{yang2017topological}. Therefore, the Berry curvature ($\Omega^{x}$, $\Omega^{y}$) does not vanish in the case of \textit{B}$\parallel$\textit{x} or \textit{B}$\parallel$\textit{y}. This leads to the large AHC observed in  Mn$_{3}$Ge as long as the magnetic field is applied along
	the \textit{x} or \textit{y} axis. Refs. \cite{tomiyoshi1983triangular,nagamiya1982triangular}
	performed a neutron diffraction experiment under a 0.8 T magnetic
	field and a change in the magnetic structure was observed depending on the direction of the applied magnetic field. Also, a
	very small field hysteresis is observed in AHE, as shown in Fig. \ref{fig:hall_hyst}, and is consistent with Ref. \cite{nayak2016large}. Therefore,
	it is very likely that just a few hundred Oe of the magnetic field
	are sufficient to change the magnetic structure from (a) to (b) of
	Fig. \ref{fig:mag_define}, which results in a very similar nature
	of Hall resistivity if the field is applied along the \textit{x}
	or \textit{y} direction, as shown in Fig. \ref{fig:hall_all_axis}.
	
	\begin{figure}
		\includegraphics[width=8cm]{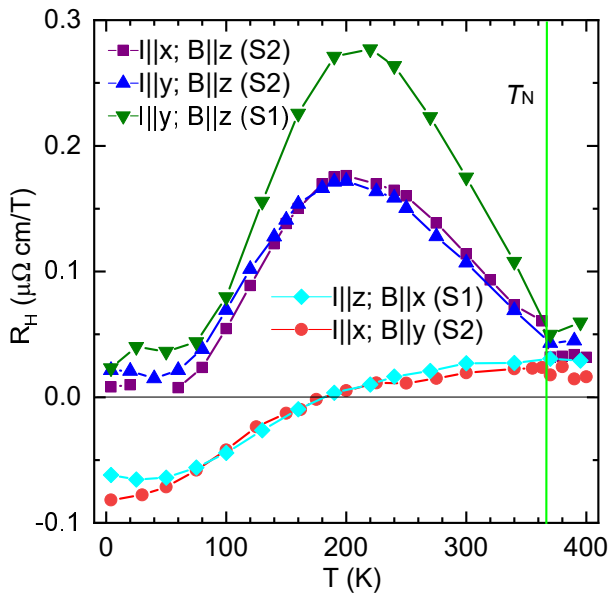} \caption{Hall coefficient determined by Hall resistivity high field slope at
			different temperatures.}
		\label{fig:hall_slope}
	\end{figure}

	The anomalous Hall conductivity (AHC) ($\sigma_{ij}^{A}$), originating
	from a non-zero Berry curvature, can be experimentally determined by:
	$\sigma_{ij}^{A}\approx-\rho_{ij}^{A}/(\rho_{ii}\rho_{jj})$. Where,
	$\rho_{ii},\rho_{jj}$ are the longitudinal resistivities at the given
	field, $\rho_{ij}^{A}$ is the Hall resistivity at \textit{B} = 0, with the field applied
	perpendicular to both the \textit{i} and \textit{j} directions. $\sigma_{ij}^{A}$
	is significantly large if the field is applied along the \textit{x} or \textit{y}
	axis. The \textit{y} axis intercept of linear fitting of Hall resistivity at high field can be regarded as $\rho_{ij}^{A}$ (example shown in Fig. \ref{fig:hall_all_axis}(a)). The temperature dependent AHC, $\sigma_{ij}^{A}(T)$, is shown in Fig. \ref{fig:cond_t}.
	We observed nearly
	the same conductivity at 4 K for S1 and S2 samples which have slightly different Mn concentrations
	- (Mn$_{3.1}$Ge and Mn$_{3.2}$Ge). Similar to Refs. \cite{chen2021anomalous,kiyohara2016giant,xu2020finite},
	the AHC was observed to be nearly constant below 150 K and 60 K for
	samples with low (S1) and high (S2) Mn concentration (Fig. \ref{fig:cond_t}), respectively.
	Hall conductivity above these temperatures decreases as the temperature
	is increased and eventually vanishes near T$_{N}$\,= 365 K. Interestingly,
	a small anomalous Hall effect was observed in the case of \textit{B}$\parallel$\textit{z}
	($\sigma_{xy}^{A}$) as well. To confirm this observation, we 
	performed a \textit{B}$\parallel$\textit{z} Hall effect experiment
	with different samples. The behavior was found to be nearly the
	same in nature but different in magnitude. In the case of \textit{B}$\parallel$\textit{z},
	small anomalous Hall and Nernst conductivity have already been reported,
	but its nature is not consistent \cite{xu2020finite,kiyohara2016giant,nayak2016large,wuttke2019berry}.
	The probable origin of the non zero $\sigma_{xy}^{A}$ could be: 
	(i) the location of Weyl points out of the \textit{a-b} plane, as theoretically
	predicted by \cite{yang2017topological}, 
	(ii) the small topological Hall effect due to very small canting of Mn moments towards the \textit{z}
	axis, as observed in \textit{M-H} with \textit{B}$\parallel$\textit{z}
	\cite{nayak2016large} and, 
	(iii) the presence of a small (2-3\%) tetragonal-Mn$_{3}$Ge impurity phase, 
	which is ferrimagnetic along the \textit{z} axis.
	However, the true origin of AHE in the case of \textit{B$\parallel$z} is still unknown.

	Using the single-band model, the carrier concentration was observed to be $\sim$ $10^{21}$/cm$^{3}$
	below 100 K when $B\parallel$\textit{z} was applied. 
	However, it is up to 10 times lower if the field is
	applied along the \textit{x} or \textit{y} axis. The Hall slopes observed
	in our samples are larger than the reported data \cite{nayak2016large,kiyohara2016giant,xu2020finite},
	which results in nearly 10 - 50 times lower carrier concentration
	compared to Ref. \cite{xu2020finite} ($\sim$ $10^{22}$/cm$^{3}$).
	The possible difference could arise because of different sample compositions
	and preparation methods.

	The Hall coefficient, ($R_{H}$), can be determined by the high field
	slope of the Hall resistivity ($R_{H}=\partial\rho_{H}/\partial B$).
	It can be seen in Fig. \ref{fig:hall_slope} that $R_{H}$ remains
	positive and shows a maximum, near 200 K, as long as the field is
	applied along the \textit{z} direction. However, it turns negative,
	near 190 K, if the field is applied along the \textit{x} or \textit{y}
	axis, irrespective of the current direction. The change in the sign of $R_H$ suggests that electrons are the dominant carrier above 190 K. However, the number of hole carriers dominates over electrons below 190 K. A similar transition in the dominant carrier concentration, at a lower temperature ($\sim$ 50 K) though, has been reported by Ref. {\cite{wang2021robust}} in the case of thin film Mn$_3$Ge. Therefore, we can argue that similar to Ref. {\cite{wang2021robust}}, a topological electronic transition occurs in single crystal Mn$_3$Ge also, where chemical potential relative to Weyl points move from the valence to the conduction band with an increase in temperature and crosses the Fermi surface near 190 K. The topological transition temperature (190 K) is very near 200 K, where the metal-semimetal transition occurs. This suggests an intricate connection between these two effects, which is expected as both the effects are determined by the electronic band structure and DOS near the Fermi surface.

	\subsection*{Planer Hall effect (PHE)}
	
\begin{figure}[h]
		\includegraphics[width=8cm]{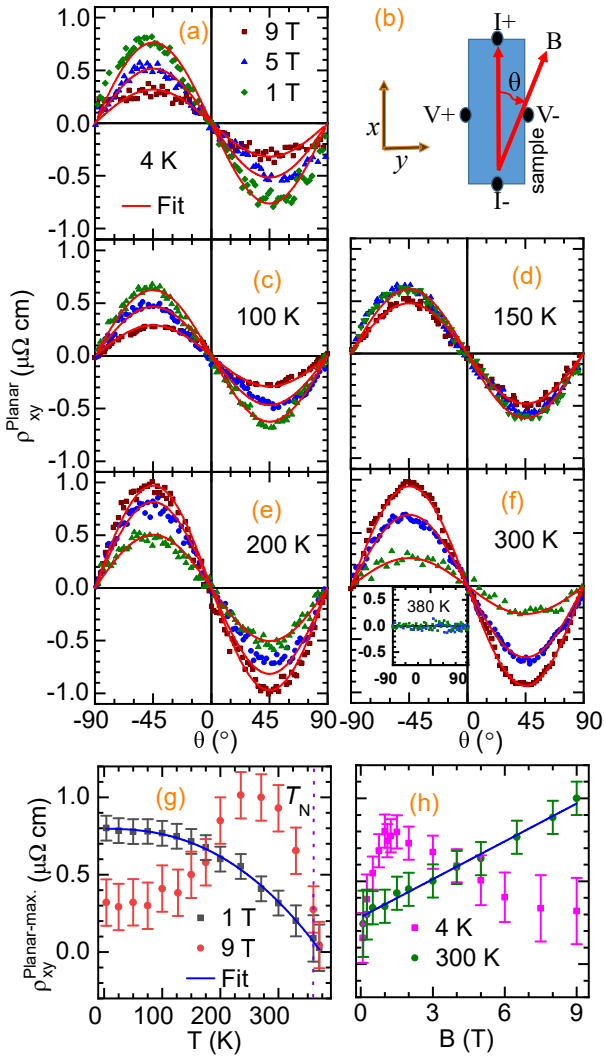} \caption{(a) shows PHE, within the \textit{x-y} plane, at 4 K. (b) illustrates the measurement setup, where the magnetic field rotates within the \textit{x-y} plane. The current and transverse voltage contacts also lie in the same plane. (c-f) show PHE at different magnetic fields applied at different temperatures mentioned in each plot. In (a-f), {\enquote*{Fit}} corresponds to the data fitting using Eqn. {\ref{eq:planer hall}}. (f) Inset: The PHE at 370 K (above $T_{\text{N}}$). (g, h) show PHE maximum magnitudes with temperature and magnetic
			field dependence, respectively. $\rho_{xy}^{Planar-max.}=\rho_{xy}^{Planar}(-45^{\circ})-\rho_{xy}^{Planar}(0)$. {\enquote*{Fit}} in (g, h) suggests fitting with (g) $a-bT^m$, (h) $B^n$, same as mentioned in the previous section. Here again, $m=2.3(1)$, and $n=1.0(1)$ was obtained.}
		\label{fig:PHEth_xy}
	\end{figure}

	Transverse resistivity can also be measured with the magnetic field rotating
	in the sample plane (as shown in Fig. \ref{fig:PHEth_xy}), which
	is known as the planar Hall effect (PHE). PHE is expected to be observed in Weyl semimetals. However, it can have non-topological origin as well {\cite{liu2019nontopological}}. Usually, the	PHE follows the following angular dependence {\cite{liu2019nontopological}}:
	\begin{eqnarray}
		\rho_{ij}^{\text{PHE}}(\theta)=-\Delta\rho_{ii}(\text{sin}\theta\text{cos}\theta)\label{eq:planer hall}; \Delta\rho_{ii}=\rho_{\bot}-\rho_{\parallel}
	\end{eqnarray}
	Where, $\rho_{ij}^{\text{PHE}}(\theta)$ is termed as the planar Hall
	resistivity with the current is applied along the \textit{i} direction
	and the voltage is measured along the \textit{j} direction, given
	that the field is rotated in the \textit{i-j} plane. It is interesting to note in the magnitude of PHE in the system is determined by the magnitude of $\theta$MR ($\Delta\rho_{ii}$).
	We have performed PHE measurement such that the magnetic field is rotated in the \textit{x-y} 
	crystallographic plane. As shown in Fig. \ref{fig:PHEth_xy},
	clear oscillations of PHE, consistent with Eqn. (\ref{eq:planer hall}),
	is observed in all the fields and temperature ranges below $T_{\text{N}}$
	(365 K). 
	
	As shown in Figs. {\ref{fig:PHEth_xy}}(g, h), the PHE magnitude at 9 T and 1 T behaves very similar to the $\theta$MR magnitude (shown in Fig. {\ref{fig:mr_th_xy}}(g, h)), and vanishes near 365 K (\textit{T}$_{\text{N}}$). The PHE magnitude at 1 T decreases monotonically with temperature, and $m=2.3(1)$ was obtained  (when fitted with $a-bT^m$),  similar to the $\theta$MR. Behavior of the low temperature (4 K) PHE magnitude is also the same as $\theta$MR, where PHE increases with field up to 1.5 T, and decreases afterward. At 300 K, PHE increases almost linearly with the magnetic field, similar to the $\theta$MR. The observations cleraly suggest that PHE magnitude follow $\theta$MR behaviour, which is expected {\cite{liu2019nontopological}}. Therefore, the origin of PHE at any temperature and magnetic field has to be the same as the origin of $\theta$MR.
	
	Further, we measured PHE in the \textit{x-z} and \textit{y-z} planes as well.
	However, a PHE oscillation was not observed in these cases. The raw data corresponding
	to different configurations are shown in the Appendix B (see Fig. \ref{fig:phe_all}). In the case of PHE measured in \textit{x-z} and \textit{y-z} planes, either the signal is absent or too weak to be analyzed. The
	absence of PHE in the case of out of \textit{a-b} plane magnetic field rotation
	may be related to the in-plane magnetic structure of the system. It
	is possible that the out-of-plane magnetic field rotation leads to equalizing the population of domains with opposite chirality, leading to neutralizing the PHE oscillations, which have a periodicity of 180$^{\circ}$. Further neutron diffraction experiments for magnetic fields out of the plane are needed to clarify the absence of PHE oscillations in these planes.
	
	\section{Single crystal neutron diffraction} \label{neutron-sec.}
	
	Magnetic Weyl semimetals primarily originate from the broken time-reversal
	symmetry in the magnetically ordered ground state.
	Therefore, slight change in magnetic structure or a sudden change in a magnetic moment with temperature could play a direct role in the observed changes of the anomalous transport properties.	Other than this, the symmetry of the magnetic structure also interferes
	with the properties observed in the magnetic Weyl semimetals, thus,
	affecting the magnitude of anomalous transport effects. As we 
	observed a significant change in the behavior of all the components of
	electrical transport effects, (even thermal transport \cite{wuttke2019berry})
	below room temperature, careful analysis of the magnetic structure
	of Mn$_{3}$Ge was required. Therefore, we performed a (unpolarized) neutron diffraction
	experiment on a single crystal of Mn$_{3}$Ge (S3) at three different temperatures
	at 300 K, 175 K, and 4 K to look for possible changes in its magnetic
	structures. The experiment was performed by selecting neutron wavelength at 0.87 {\AA}. The data analysis was performed using JANA2006 software
	\cite{JANA2006}. Since propagation vector, \textbf{k} = 0 for Mn$_{3}$Ge
	\cite{tomiyoshi1983triangular,soh2020ground}, the nuclear and magnetic
	reflections coincide. Therefore, high \textit{Q} ($0.7>\text{sin}\theta/\lambda>\thinspace$0.4)
	reflections \footnote{Due to the specific form of magnetic form factor we expect a negligible contribution of magnetic intensities at high Q.} were fitted to obtain the nuclear parameters, which were in agreement
	with the results from X-ray diffraction of the same sample (S3). The Mn atoms occupying Ge sites were assumed to be non-magnetic while performing the
	magnetic refinement. 
	
\begin{figure}
		\includegraphics[width=8cm]{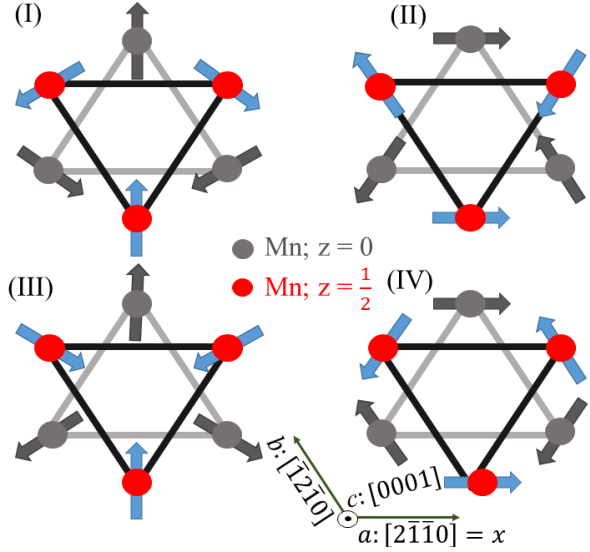} \caption{Different magnetic structures which are reasonably based on the magnetization data and allowed by P$6_{3}$/\textit{mmc} space group. In (I, III) moments are lying along the {[}$01\bar10${]} , and in (II, IV) moments are parallel to the {[}$2\bar1\bar10${]} direction.}
		\label{fig:mag_structure}
	\end{figure}

	The representation analysis of space group P$6_{3}$/\textit{mmc}
	for \textbf{k} = 0 gives 18 basis vectors, corresponding to 18 types
	of possible magnetic structures. Most of them could be rejected based
	on the magnetization data, except for four magnetic structures shown in
	Fig. \ref{fig:mag_structure}. The fitting parameter ($\chi^2$) corresponding to these magnetic structures is mentioned in Table \ref{structural_parameters_2} in the appendix. There are three (120$^{\circ}$)
	magnetic domains possible for each structure in Fig. \ref{fig:mag_structure}.
	Therefore, unequal populations of the domains were also taken into consideration,
	and the refinement was performed by fitting data with different magnetic
	structures at all three temperatures. We observed that the neutron	diffraction data at all three temperatures (4 K, 175 K, 300 K) fit best with magnetic structures I and II (Fig. \ref{fig:mag_structure}) only, as mentioned in Table \ref{structural_parameters_2} in the Appendix C. Since magnetic structure factors for I and II are same for all the reflections, even if unequal domain population is taken into account, they cannot be differentiated using unpolarized single crystal neutron diffraction. However, Ref. \cite{soh2020ground} has reported Fig. \ref*{fig:mag_structure} (I) to be the ground state magnetic structure of Mn$_3$Ge. Therefore, we also considered magnetic structure I as the ground state magnetic structure of Mn$_3$Ge.
	
	\begin{figure}[h]
		\includegraphics[width=8cm]{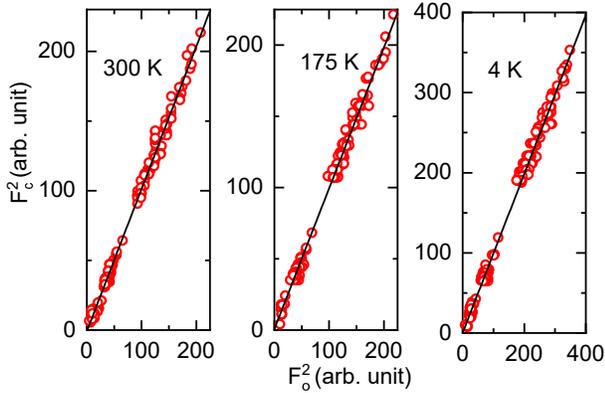} \caption{Observed integrated intensities compared with the calculated intensity,
			corresponding to magnetic structure - I from Fig. \ref{fig:mag_structure},
			at different temperatures. Since \textit{Q} = 0, red dots represent the
			the sum of nuclear and magnetic reflections. Here, F$_{o}^{2}$, F$_{c}^{2}$
			corresponds to observed and calculated intensities, respectively.}
		\label{fig:neutron_all_t}
	\end{figure}
	
	\begin{figure}[h]
		\includegraphics[width=8cm]{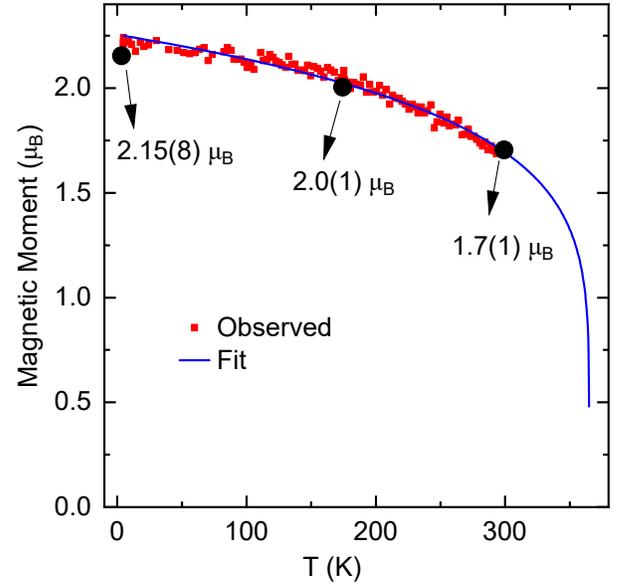} \caption{Variation of the magnetic moment with temperature,
			determined by the (101) peak. The data follow the $\mu(T)=\mu_0(1-T/T_{\text{N}})^{\beta}$ type behavior with $\beta=0.16(1)$. Black dots indicate the magnetic moment determined from the refinement of the data at 4 K, 175 K, and 300 K.}
		\label{fig:neutron_t_down}
	\end{figure}
	
	\begin{figure}
		\includegraphics[width=8cm]{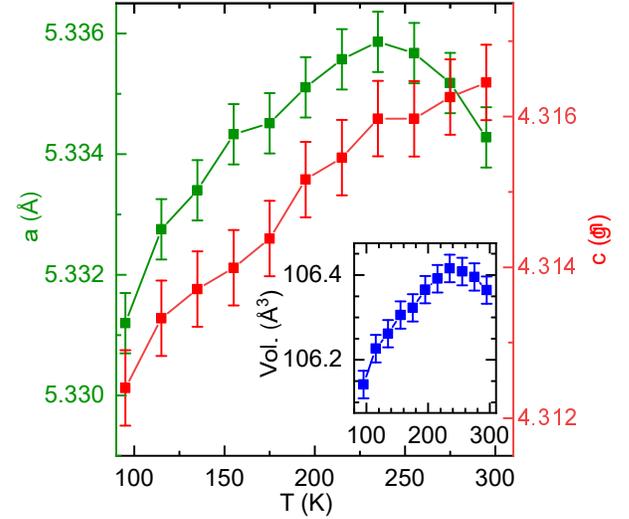} \caption{Variation of lattice parameters of Mn$_{3}$Ge (S2) at different temperatures.
			Inset: Temperature dependence of the lattice volume.}
		\label{fig:xrd_t}
	\end{figure}
	
	\begin{table}
		\caption{Results of the refinement of single crystal neutron diffraction performed
			at different temperatures. Here, $\chi^2$, R, R$^w$ are the fitting parameters provided by Jana2006 software. U$_{\text{iso}}$ defines the isotropic thermal parameter of the corresponding element.}

		\begin{ruledtabular} %
			\begin{tabular}{cccc}
				Temperature & \multirow{1}{*}{4 K} & 175 K & 300 K\tabularnewline[\doublerulesep]
				\hline 
				\noalign{\vskip\doublerulesep}
				Moment & 2.15(8) & 2.01(7) & 1.68(9)\tabularnewline
				$\chi^{2}$ & 1.82 & 1.35 & 1.79\tabularnewline
				R & 2.76 & 2.32 & 2.37\tabularnewline[\doublerulesep]
				\multicolumn{1}{c}{R$^{w}$} & 3.15 & 2.53 & 3.22\tabularnewline[\doublerulesep]
				U$_{\text{iso}}${[}Mn{]} & 0.0036(2) & 0.0058(1) & 0.0082(3)\tabularnewline[\doublerulesep]
				U$_{\text{iso}}${[}Ge{]} & 0.0014(2) & 0.0031(2) & 0.0050(2)\tabularnewline[\doublerulesep]
				Domain ratio & 55:35:10 & 50:35:15 & 52:42:6\tabularnewline[\doublerulesep]
				No. of reflections & 229 & 165 & 175\tabularnewline[\doublerulesep]
		\end{tabular}\end{ruledtabular} \label{structural_parameters_1}
	\end{table}
	
	Calculated vs. observed intensities, corresponding to the magnetic structure
	I, are shown in Fig. \ref{fig:neutron_all_t}, and the fitting parameters
	at each temperature are given in Table \ref{structural_parameters_1}. Based on the analysis, it
	can be concluded that no change in the magnetic structure was observed
	starting from 300 K to 4 K. The temperature-dependent magnetic moment
	was also determined by collecting the strongest magnetic peak (1 0 1) at
	several temperatures. As shown in Fig. \ref{fig:neutron_t_down}, the magnetic moment increases smoothly with decrease in	temperature. The magnetic moment can be fitted with a power law of the form $\mu(T)=\mu_0(1-T/T_{\text{N}})^{\beta}$, 
	where $\mu_0$ is the magnetic moment at $T=0$ K, and
	found to be 2.25(1) $\mu_{\text{B}}$. $\beta$, 
	which is an exponent, was found to the 0.16(1). The $T_{\text{N}}$ = 165 K was fixed.
	
	Our neutron diffraction analysis at zero magnetic fields, along with
	field-dependent studies by Refs. \cite{nagamiya1982triangular,tomiyoshi1983triangular},
	suggest that the magnetic structure remains the same below 300 K. Therefore, the change in the behavior of transport properties is
	not directly linked to the magnetic structure or the ordered magnetic
	moment. To look further,
	we measured lattice parameters as a function of temperature.
	Fig. \ref{fig:xrd_t} shows the temperature variation of the lattice parameters.
	The in-plane lattice parameter, \textit{a}, shows a maximum near
	230 K, whereas the \emph{c} lattice parameter varies smoothly. Negative
	thermal expansion was observed in the temperature range of 300 K to 230
	K, same as reported by \cite{song2018opposite,sukhanov2018gradual}.
	Below 230 K, it follows positive thermal expansion as expected for
	any normal metal. Since transport properties show a change in the
	behavior near or below 200 K, it is very likely that the transport
	properties are sensitive to the changes in the lattice parameters.
	
	\section{Conclusion}
	
	An increase in the excess Mn concentration ($\delta$) in Mn$_{3+\delta}$Ge increases the magnitude of resistivity, thus conductivity decreases.	However, the intrinsic nature of the transport behavior of Mn$_3$Ge (S2, S3) remains unchanged. The LMR  shows sharp negative behavior at low fields, which turns towards positive beyond 1.5 T. Therefore, the presence of a chiral anomaly in Mn$_3$Ge cannot be justified. At the high field, all the electrical transport effects show a change in behavior near 200 K, where metal-semimetal transition occurs. Analysis of transport measurements shows that the high field LMR and $\theta$MR of Mn$_3$Ge are possibly driven metallic or semimetallic nature of the sample, below and above 200 K, respectively. Dominant carrier type also changes from hole to electrons near 190 K. Since electrical transport effects are determined by the band structure near the Fermi surface, an electronic topological transition is very likely to be present in the Mn$_3$Ge near 200 K. Neutron diffraction measurements show that the magnetic structure remains same below and above 200 K. However, in-plane lattice parameter shows an unusual behavior near 230 K. It is possible that the change in electrical properties, near 200 K, could possibly be driven by the change in lattice parameter of Mn$_3$Ge.
	
	\section{Acknowledgements}
	Single crystal neutron diffraction data used in this publication were measured at the instrument HEiDi jointly operated by the Institute of Crystallography of RWTH Aachen University and the J\"ulich Centre of Neutron Science, Forschungszentrum J\"ulich GmbH within the JARA cooperation at the Heinz Maier-Leibnitz Zentrum (MLZ). 
	
	\global\long\def\appendixname{APPENDIX}%

	\appendix
	
	\section{EXPERIMENTAL DETAILS}
	
\begin{figure}[H]
		\begin{center}
			\includegraphics[width=7cm]{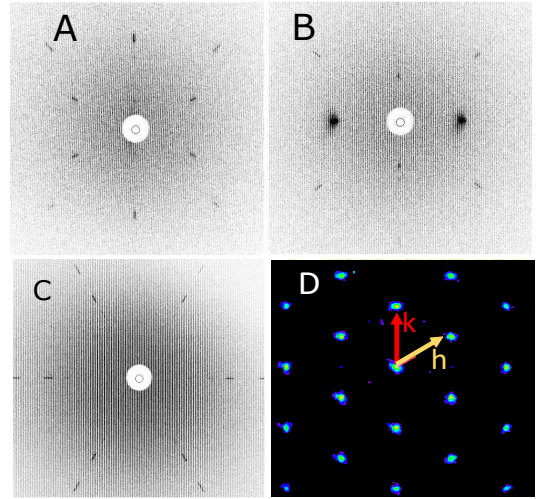}
			\caption{Laue diffraction pattern obtained from the single crystals of Mn$_{3}$Ge
				when X-ray beam was parallel to (A) \textit{x} {[}$2\bar1\bar10${]},
				(B) \textit{y} {[}$01\bar10${]}, (C) \textit{z} {[}0001{]} axes.
				(D) Shows single crystal x-ray diffraction pattern in (\textit{h k} 0) plane at 300 K}
			\label{fig:laue}
		\end{center}
	\end{figure}

	\textbf{X-Ray diffraction}: High-intensity Laue spots were also observed in the Laue diffraction
	of the single crystals. The Laue diffraction pattern for X-ray parallel
	to the \textit{x, y, z} crystallographic axis are shown in Fig. \ref{fig:laue}.
	Single crystal X-ray diffraction was performed using SuperNova (Rigaku Oxford Diffraction) instrument. High-intensity reflections were observed, as shown in \ref{fig:laue}(D), which suggests the high quality of
	the single crystals.

	\begin{figure}[h]
		\includegraphics[width=7cm]{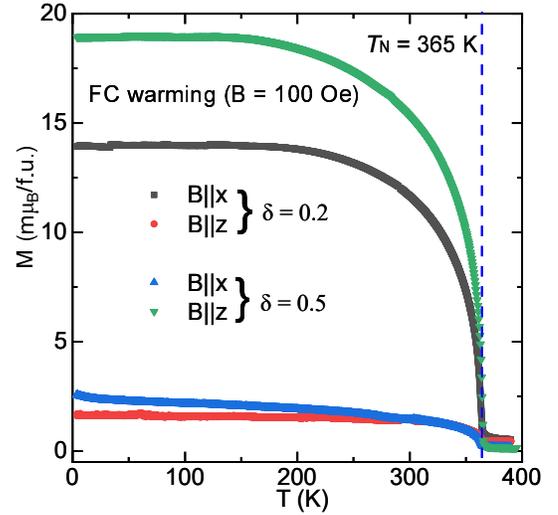} \caption{Magnetization of two different samples with low ($\delta=0.2$) and
			high ($\delta=0.5$) Mn concentration. Magnetization was measured
			during warming in field-cooled (FC) condition with 100 Oe magnetic
			field applied in plane (\textit{x} axis) and out of plane (\textit{z}
			axis).}
		\label{fig:mt}
	\end{figure}
	
	\begin{figure}[h]
		\includegraphics[width=8cm]{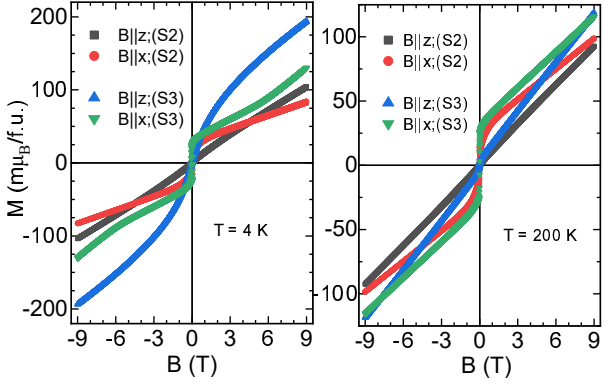} \caption{\emph{M-H} data for different samples with the magnetic field applied
			in the plane (\textit{x} axis) and out of the plane (\textit{z} axis).}
		\label{fig:mh}
	\end{figure}
	
	\textbf{Magnetization}: The magnetization of all the single crystals of Mn$_{3}$Ge was measured
	using the Quantum Design PPMS setup, with the magnetic field applied
	along the $x$, $y$, and $z$ directions. The N\'eel temperature (\textit{T}$_{\textup{{N}}}$)
	was observed to be near 365 K for all the samples. For both the samples, magnetization starts to saturate below 200 K. Magnetization behavior was found to be very similar to each other for S1 and S2 samples.
	However, a slightly different magnetization behavior was observed, below
	25 K, in the case of the S3 sample, when the field was applied along \textit{\emph{the}}\textit{
		z} direction. M-T, M-H for S2 and S3 samples are compared in Fig.
	\ref{fig:mt} and Fig. \ref{fig:mh}. The magnetization data (for S2) shows
	residual magnetic moment of $\sim$ 20 m$\mu_{B}$/f.u. and 4
	m$\mu_{B}$/f.u. when field was applied in the \textit{a - b} plane,
	and along the \textit{c} axis, respectively. This confirms that
	along with dominant antiferromagnetic (AFM) structure, Mn moments
	are canted ferromagnetically in the \emph{a-b} plane \cite{nayak2016large,kiyohara2016giant}.
	Since a small amount of tetragonal phase (2-5\%) was observed in all
	the samples, its effect on magnetization data was important to determine.
	\emph{M-H} at 390 K, (paramagnetic regime for hexagonal phase), was
	performed. We observed only paramagnetic behavior even though the
	tetragonal phase is ferrimagnetic. Therefore, we can conclude that
	a small (tetragonal) impurity phase has a negligible effect on the
	magnetization and electrical transport measurement.
	
	\section{TRANSPORT MEASUREMENT}
	\label{appendix:transport}
\begin{figure}[h!]
		\includegraphics[width=8.5cm]{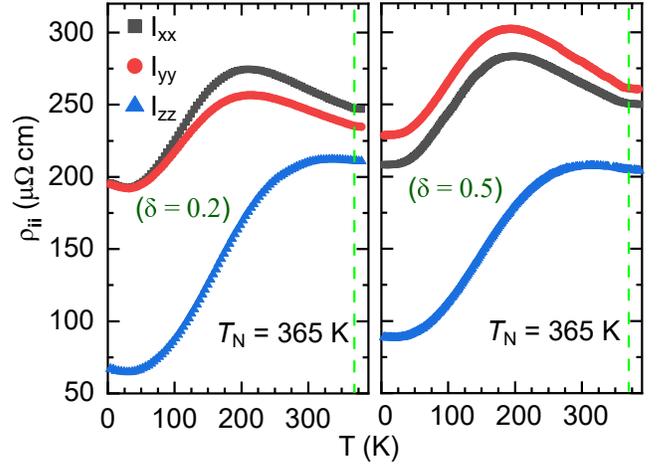} \caption{Longitudinal resistivity of Mn$_{3+\delta}$Ge single crystal along
			the different crystallographic directions. $\delta=0.2, 0.5$ correspond to the S2 and S3 samples, respectively.}
		\label{fig:rt}
	\end{figure}

	Longitudinal resistivity corresponding to the two samples with significantly different Mn concentrations is shown in Fig. \ref{fig:rt}. The resistivity along \textit{x, y} axis looks very similar to each other. However, it shows significantly different behavior along the \textit{z} axis.
	\subsection*{Current jetting effect}
\begin{figure}[h!]
		\includegraphics[width=8cm]{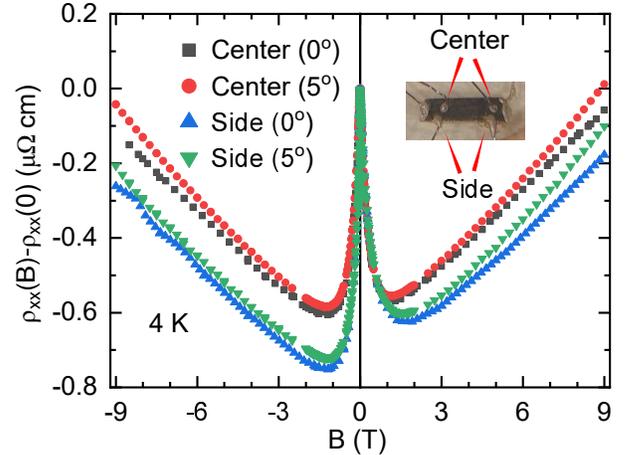}\\
		\caption{LMR of Mn$_{3}$Ge (S1) at 4 K. \enquote*{Center} and \enquote*{Side} terms denote
			the contact positions are shown in the inset. 0$^{\circ}$ and 5$^{\circ}$
			denotes the angle between the electric current and the magnetic field.}
		\label{fig:mr_side}
	\end{figure}

	The current jetting effect can also be the possible reason behind
	the negative MR observed in samples, however, we can nullify this effect
	based on three arguments: (1) We observed a very similar nature
	of MR for all three samples. (2) samples were carefully cut to create
	a parallel electric field region. (3) voltage was measured at different
	positions of the sample to see the effect of a non-uniform electric
	field if present. The behavior of LMR along \textit{x} axis has a
	negligible effect of voltage contact positions - either at the center
	or sides, as shown in Fig. \ref{fig:mr_side}. Since the magnitude
	of MR is nearly the same in all directions, the current jetting effect
	cannot be expected to be significant.
	
	\begin{figure}[h!]
		\includegraphics[width=8cm]{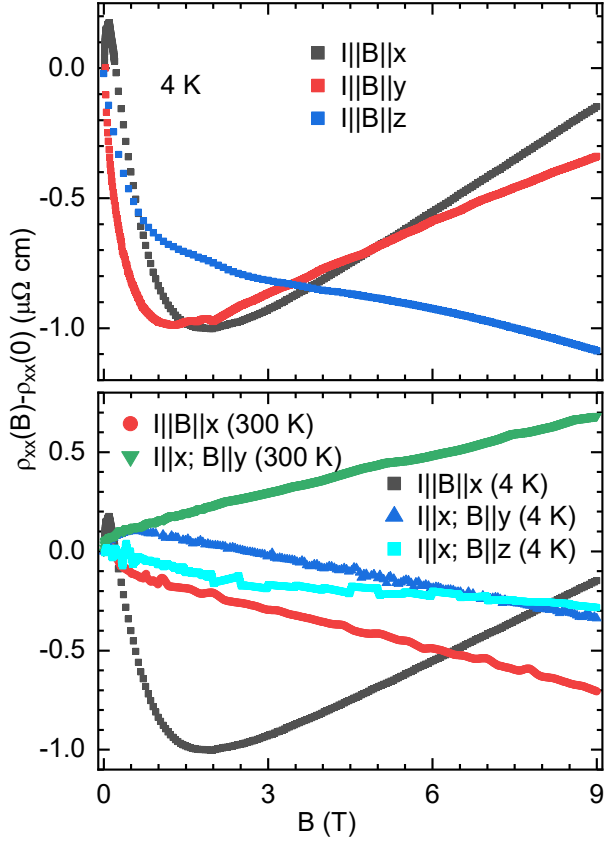} \caption{Top: LMR of Mn$_{3}$Ge along different axis. Bottom: MR of the sample
			at 4 K and 300 K, with $B$$\perp$\textit{I} configuration. (\textit{I}
			= electric current).}
		\label{fig:mr_para_perp}
	\end{figure}
	
	\begin{figure}[h!]
		\includegraphics[width=8cm]{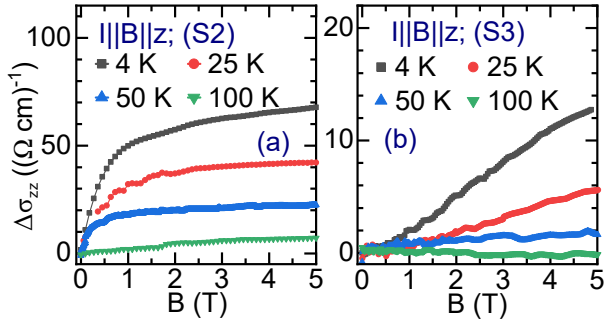}\\
		\caption{LMC along \textit{z} axis (\textit{B}$\parallel$\textit{I}$\parallel$\textit{z}) for S2 and S3 samples are shown in (a) and (b), respectively.}
		\label{fig:LMC_B_IzzBz}
	\end{figure}
	
	\begin{figure}[h!]
		\includegraphics[width=8.2cm]{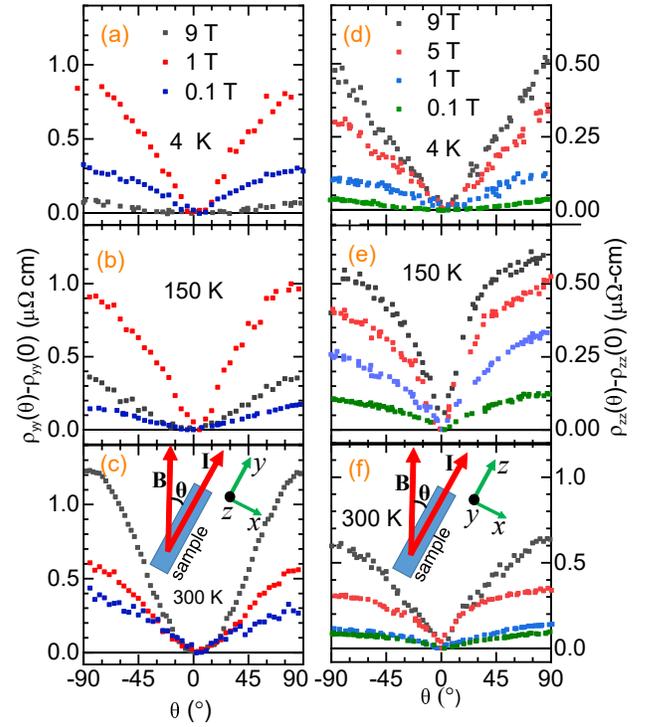} \caption{Angle dependent MR at different magnetic field and temperature. (a-c):
			The electric current is applied along the \textit{y} axis, and sample
			is rotated such that \textit{B} rotates from the \textit{y} axis towards the \textit{x} axis. (d-f): The electric current is applied along the \textit{z} axis, and the sample
			is rotated such that \textit{B} rotates from the \textit{z} axis towards the \textit{x} axis.}
		\label{fig:mr_th_yx_zx}
	\end{figure}

	\subsection*{$\theta$MR (\textit{z-x} plane rotation)}
	
\begin{figure}[h!]
		\includegraphics[width=8.7cm]{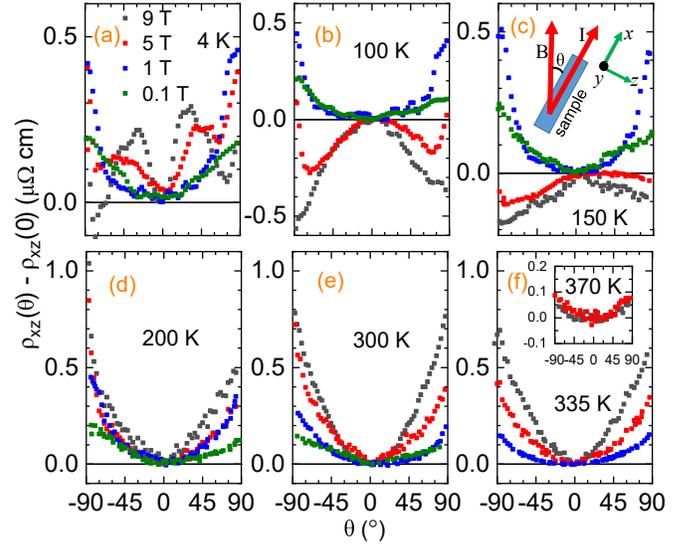} \caption{Angle dependent MR of Mn$_{3}$Ge (S1) at different magnetic fields
			and temperatures. The electric current is applied along \textit{x}
			axis and sample are rotated towards \textit{z} axis.}
		\label{fig:mr_th_xz}
	\end{figure}

	As shown in Figs. (\ref{fig:mr_th_yx_zx}(d-f), \ref{fig:mr_th_xz}), positive $\theta$MR is observed in the case of \textit{I}$\parallel$\textit{B}$\parallel$\textit{z}; \textit{B} rotation from $z$ $\rightarrow$$x$, and \textit{I}$\parallel$\textit{B}$\parallel$\textit{x}; \textit{B} rotation from $x$ $\rightarrow$$z$. In the case of \textit{I}$\parallel$\textit{B}$\parallel$\textit{z}; \textit{B} rotation from $z$ $\rightarrow$$x$, $\theta$MR increases (decreases) with field (temperature) at all the temperatures (field). However, the magnitude of $\theta$MR is non-monotonic with field and temperature for \textit{I}$\parallel$\textit{B}$\parallel$\textit{x}; \textit{B} rotation from $x$ $\rightarrow$$z$.  Since Mn$_3$Ge has magnetic anisotropy in \textit{z-x} plane, the observed $\theta$MR may result from the magnetization of the sample.
	
	\begin{figure}[h!] 
		\includegraphics[width=5cm]{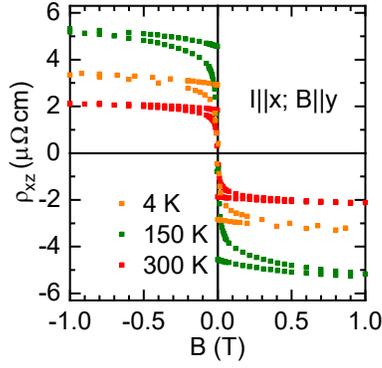} \caption{Hall resistivity of Mn$_3$Ge at low magnetic field. The magnetic field was applied along the \textit{y} axis. }
		\label{fig:hall_hyst}
	\end{figure}
	
	\begin{figure}[h!]
		\includegraphics[width=8.5cm]{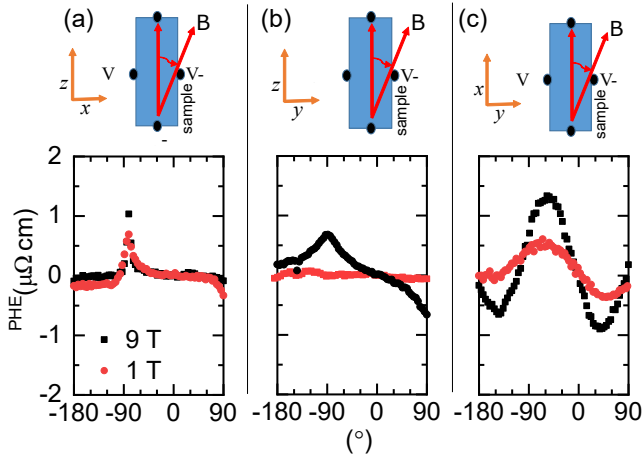} \caption{Raw data of PHE corresponding to out of plane rotation (a, b) and
			in-plane (c) rotation of the magnetic field at \textit{T} = 300 K.
			Oscillations are observed in the case of \textit{x-y} plane rotation
			only, which is shown in (c). Signals observed in the other two cases are
			most likely coming from the normal Hall effect due to slight misalignment
			of the sample or magnetic field.}
		\label{fig:phe_all}
	\end{figure}

	\begin{table} [h!]
		\caption{Results of the refinement of single crystal neutron diffraction performed
			at different temperatures, for four different types of magnetic structures shown in Fig. \ref{fig:mag_structure}. Here, $\chi^2$ is the goodness of fitting parameters provided by Jana2006 software.}

		\begin{ruledtabular} %
			\begin{tabular}{cccc}
				Temperature & \multirow{1}{*}{4 K} & 175 K & 300 K\tabularnewline[\doublerulesep]
				\hline 
				\noalign{\vskip\doublerulesep}
				$\chi^{2}$ (model I) & 1.82 & 1.35 & 1.79\tabularnewline
				$\chi^{2}$ (model II) & 1.82 & 1.35 & 1.79
				\tabularnewline
				$\chi^{2}$ (model III) & 4.18 & 3.43 &  2.58
				\tabularnewline
				$\chi^{2}$ (model IV) & 3.95 & 3.31 &  2.4
				
				\tabularnewline[\doublerulesep]
		\end{tabular}\end{ruledtabular} \label{structural_parameters_2}
	\end{table}
	
	\section{Neutron diffraction }
	The neutron diffraction analysis at different temperatures was performed using various models mentioned in the main text. The fitting parameter corresponding to different models is shown in Table \ref{structural_parameters_2}.
	
	\vspace{2cm}
	
	\bibliographystyle{apsrev}
	\bibliography{Mn3Ge_reference}

\end{document}